\documentclass[reprint,showpacs,amsfonts,amsmath,amssymb,prb]{revtex4-1}
\usepackage{graphicx,dcolumn,amsmath,amstext}

\usepackage{bm}
\begin{document}

\title{Fermi arcs, pseudogap and collective excitations in doped Sr$_2$IrO$_4$:\\ A generalized fluctuation exchange study}

\author{Hu Wang}
\author{Shun-Li Yu}
\author{Jian-Xin Li}
\affiliation{National Laboratory of Solid State Microstructures
and Department of Physics, Nanjing University, Nanjing 210093,
China\\
Collaborative Innovation Center of Advanced Microstructures,
Nanjing University, Nanjing, China}

\date{\today}

\begin{abstract}
Motivated by recent experimental measurements,
we study the quasiparticle spectra and the collective excitations in doped  Sr$_2$IrO$_4$,  in which the
interesting interplay between the electronic correlations and strong spin-orbital coupling (SOC) exists.
To include the SOC,  we use the Hugenholtz  diagrams to extend the
fluctuation exchange (FLEX) approach to the case where the
SU(2) symmetry can be broken. By using this generalized FLEX
method, we find a weak pseudogap behavior near $(\pi,0)$ in the
slightly electron-doped system, with the corresponding Fermi arc
formed by the partial destruction of  Fermi surface. Similar
features also appear in the hole-doped system, however, the
position of the Fermi arc is rotated $45^\circ$ with respect to
the former. These results are consistent with the recent
angle-resolved photoemission spectra in Sr$_2$IrO$_4$. We
elaborate that these anomalous phenomena are caused by the scatterings of quasiparticles off
the isospin fluctuation derived from the effective
$J_{\text{eff}}=1/2$ doublet.
\end{abstract}


\maketitle

\section{introduction}

Recently, the $5d$ transition-metal iridium oxides have attracted
significant attention, because they exhibit a number of exotic
phenomena induced by the spin-orbital coupling (SOC) and the
correlation effects of electrons~\cite{review,summary}. Of these
iridates, Sr$_2$IrO$_4$ of particular interest for it shares many analogies with the parent
compound La$_2$CuO$_4$ of high-$T_c$ cuprate superconductors, such
as the same layered perovskite crystal structure \cite{finding1},
the same antiferromagnetic (AFM) insulating ground
state~\cite{finding2}, the similar magnetic excitation
spectrum~\cite{magnon} and electronic structure~\cite{waterfall}.
It has also been proposed theoretically to realize the unconventional
superconductivity via doping~\cite{SC_summary}.

In particular, the recent
angle-resolved-photoemission-spectroscopy (ARPES) measurements on
the slightly electron-doped Sr$_2$IrO$_4$  show a
temperature-dependent pseudogap phenomenon~\cite{arc1}. The
intensity of the spectra is much suppressed in an extended region
near $(0,\pi)$, resulting in the Fermi arc which resembles the
case in the hole-doped cuprates~\cite{pseudogap_cuprates}. For the
slightly hole-doped Sr$_2$IrO$_4$, another ARPES experiment also
exhibits the trait of pseudogap~\cite{YCao}, but with the
suppression of spectral near the $(\pi/2,\pi/2)$ point. As the
pseudogap puzzle is the longstanding unsolved problem in cuprate
superconductors, studying  the origin of the Fermi arcs and
pseudogaps in doped Sr$_2$IrO$_4$ is not only important for this
material itself,  but might also  help to investigate the similar
phenomena in  high-$T_c$ cuprates and/or other materials.

Sr$_2$IrO$_4$  shows a multi-orbital electronic structure  where
the $t_{2g}$ and $e_g$ orbitals are separeted  by large crystal
field. The five electrons (one hole) reside in the lower $t_{2g}$
manifold of $xy,xz,yz$ orbitals. In spite of the large band width
and small Coulomb interactions, Sr$_2$IrO$_4$ is an
antiferromagnetic insulator~\cite{finding1,finding2}. It has been
proposed that the SOC breaks this sixfold degenerate manifold into
completely filled $J_{\text{eff}}=3/2$ bands and a half-filled
$J_{\text{eff}}=1/2$ band (Kramers doublet) which is further split
by the relatively small Coulomb interactions. Thus, it can be
simplified to an effective one-band half-filled system, which
hosts an isospin $J_{\text{eff}}=1/2$ spin-orbital Mott insulating
ground state~\cite{Kim1,science_kim}. However, the validity of
this isospin $J_{\text{eff}}=1/2$ Mott picture including its
robustness to dopings still remains an open
question~\cite{Hsieh,Arita,liu,Li}.

Motivated by these experimental and theoretical progress, we carry out a theoretical
study of the collective excitations and
quasiparticle spectra in the doped Sr$_2$IrO$_4$, based on the
three-orbital Hubbard model with the inclusion of the SOC.

The multi-orbital structure together with the SOC confines the
exact diagonalization and quantum Monte Carlo (QMC) methods to
very small systems. In view of this, the fluctuation-exchange
(FLEX) approximation is a good
alternative~\cite{flex-1,flex-3,Takimoto}. The FLEX method has
advantages to handle various collective fluctuations, and  the
calculation results agree well with the QMC simulations for the
Hubbard model with the moderate on-site interaction
$U$~\cite{flex-1,flex-3}. Up to now, the FLEX approach has been
extensively applied to the high-$T_c$ cuprates~\cite{flex_Cu}, the
iron-based superconductors~\cite{flex_Fe}, and other correlated
electron systems~\cite{flex_other,HuWang}. However, the previous
FLEX calculations are restricted to the case with the spin
rotational invariance. In order to include the SOC, we will use
the technique of Hugenholtz diagrams to extend the FLEX approach
to more general cases, where the SU(2) symmetry can be broken.

Based on this method, we find that  the spectral function of
quasiparticles is much suppressed at parts of momenta in the
lightly doped region, suggesting the emergence of the pseudogap.
Explicitly, the  suppression occurs near $(0,\pi)$ point for
slightly electron doping and thus leads to the Fermi arc near
$(\pi/2,\pi/2)$, while their positions are reversed in the
hole-doped side. These results are consistent with the recent
ARPES observations~\cite{arc1,YCao}. We elaborate that the Fermi
arcs and pseudogaps are mainly caused  by the isospin fluctuation
with momentum ($\pi,\pi$), which overwhelms both the spin and
orbital fluctuations. We have also studied the evolution of
various collective fluctuations with doping and find that the
isospin fluctuation dominates in the  region from 30\% hole-doping
to 50\% electron-doping. These results suggest that the scenario of
$J_{\text{eff}}=1/2$ spin-orbital Mott insulator is applicable to
the parent and the extensively doped compounds of Sr$_2$IrO$_4$.

\section{Model and method}
\subsection{Three orbital Hubbard model}
We begin with the  $t_{2g}$ three-orbital Hubbard model on the
square lattice~\cite{three1}: $H\!=\!H_{kin}\!+\!H_{SOC}+\!H_I$.
The kinetic and SOC Hamiltonians read
\begin{equation}
\begin{split}
  H_{kin} &= \sum_{\bm km\alpha}\epsilon_m(\bm k)d^\dag_{\bm{k}m\alpha}d_{\bm{k}m\alpha},\\
  H_{SOC} &= \sum_{\bm kmn\alpha\beta}\frac{\xi_{SOC}}{2}{\bm L}_{mn} \cdot {\bm\sigma}_{\alpha\beta} d^\dag_{\bm{k}m\alpha}d_{\bm{k}n\beta},
\end{split}
{\label{eq:H0}}
\end{equation}
where  $d^\dag_{\bm{k}m\alpha}$ ($d_{\bm{k}m\alpha}$) creates
(annihilates) a $m$-orbital electron with spin $\alpha$  and
momentum $\bm k$. $\xi_{SOC}$ denotes the magnitude of SOC, and
${\bm L}$ and ${\bm\sigma}$ are the  orbital angular momentum
operator and Pauli matrix. Explicitly, the nonzero elements of
${\bm L= (L^x,L^y,L^z)}$ for  $yz$(1), $zx$(2), and $xy$(3)
orbitals are
$L^x_{23}=-L^x_{32}=L^y_{31}=-L^y_{13}=L^z_{12}=-L^z_{21}=i$.
 The single-particle dispersions are given by 
$\epsilon_1(\bm k)=-2t_5\cos{k_x}-2t_4\cos{k_y}$, $\epsilon_2(\bm
k)=-2t_4\cos{k_x}-2t_5\cos{k_y}$, and $\epsilon_3(\bm
k)=-2t_1(\cos{k_x}+\cos{k_y})
-4t_2\cos{k_x}\cos{k_y}-2t_3(\cos{2k_x}+\cos{2k_y}) +\mu_3$, with
parameters
$(t_2,t_3,t_4,t_5,\mu_3)=(0.5,0.25,1.03,0.17,-1.0)t_1$~\cite{three1}.
Hereafter, we set $\xi_{SOC}=1.03\,t_1$ without annotation, and  $t_1=1$ as the energy unit. The interaction part
on the  $l$-site is given by
\begin{equation}
\begin{split}
H_I(l)=
&\frac{1}{2}\sum_{ijmn}\sum_{\alpha\beta\mu\nu}\delta_{\alpha\nu}\delta_{\beta\mu}
\{U\delta_{i=j=m=n} (1-\delta_{\alpha\beta}) \\
&+ U^{\prime}\delta_{in}\delta_{jm}(1-\delta_{ij}) + J\delta_{im}\delta_{jn}(1-\delta_{ij})\\
&+ J^{\prime}\delta_{ij}\delta_{mn}(1-\delta_{im})(1-\delta_{\alpha\beta})\}
d^\dag_{li\alpha}d^\dag_{lj\beta}d_{lm\mu}d_{ln\nu},
\end{split}
{\label{eq:Hi}}
\end{equation}
where $U$ ($U^{\prime}$) is the intra-orbital (inter-orbital)
Coulomb interaction, $J$ the Hund's coupling and $J^{\prime}$ the
inter-orbital pair hopping. As usual, we set $J^{\prime}=J$ and
use the relation $U=U^{\prime}+2J$. Note that we have fabricated
$H_I$ to be a compact form to conveniently construct the
Hugenholtz vertices.

\subsection{Generalized FLEX method}

In this section, we give a thorough introduction of the generalized FLEX
method, which can naturally include the SOC term. The FLEX approach
originates from the conserving approximation theory proposed by
Baym and Kadanoff~\cite{Baym}. In this formulism, the
closed-linked $\Phi$ diagrams (also known as Luttinger-Ward
functional~\cite{Luttinger}) yield the self-energy and the
irreducible interaction vertices in a thermodynamic
self-consistent manner, in which the conservation laws on the
particle number, momentum, angular momentum and energy are
respected. The FLEX method pioneered by Bickers and Scalapino
~\cite{flex-1} is  the simplest application of the
Baym-Kadanoff formulism beyond the Hartree-Fock level, and has
been widely applied to the single- and multi-orbital Hubbard
models~\cite{Takimoto}, in which the spin rotational invariance is respected.
When the SU(2) spin symmetry is conserved,  the
scattering processes explicitly include the equal-spin
particle-particle (PP), opposite-spin PP, equal-spin particle-hole
(PH), and opposite-spin PH channels (see
Ref.~\cite{flex-1}). If the  SU(2) symmetry is broken, however, one
must consider the PP and PH fluctuations in a comprehensive manner
for there are  mixtures of equal-spin and opposite-spin scatterings.

\begin{figure}
  \centering
  \includegraphics[scale=1.3]{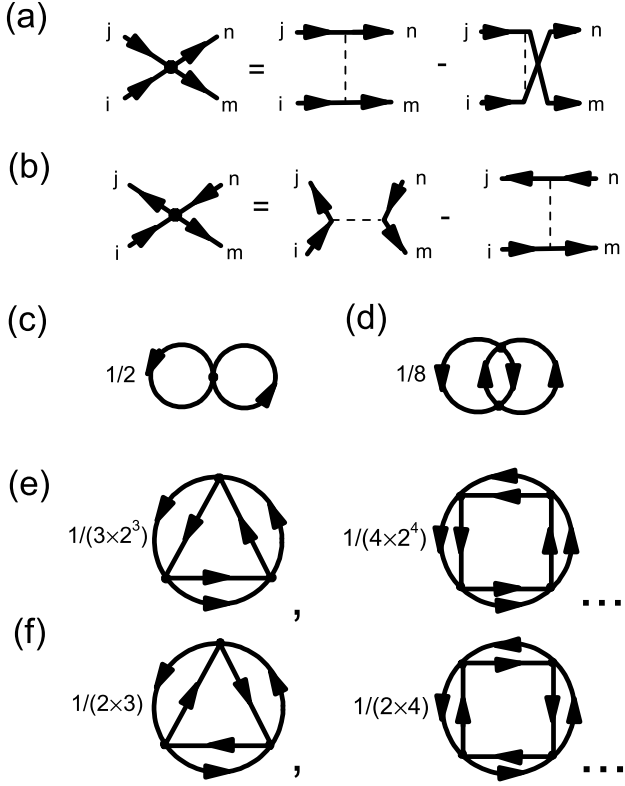}
  \caption{ (a) and (b) are bare Hugenholtz vertices for the PP and PH channels.
  (c) and (d) represent the HF term and the second order term.
  (e) and (f) are  RPA forms of the closed linked $\Phi$ diagrams for the PP and PH channels.
  The numbers in front of  diagrams (c)-(f) are the symmetry factors.
  Note that there are no anomalous Green's function lines because we only  consider the normal state properties here.
  }
  {\label{phi}}
\end{figure}

We  employ the Hugenholtz diagrams~\cite{Negele} to  extend this
method to  the SU(2) broken cases. Considering the on-site
two-body potential
\begin{equation}
 \frac{1}{2}\sum_{ijmn} \langle ij|V|mn\rangle c^\dag_{i}c^\dag_{j}c_{n}c_{m},
 \label{eq:twobody_vertex}
\end{equation}
where the index $i$  denote  both the spin and orbital degrees of
freedom. The Hugenholtz bare vertices
for the PP and PH channels  are defined as
\begin{equation}
\begin{split}
 \Gamma^p_{ij,mn} &\equiv \langle ij|V|mn \rangle-\langle ij|V|nm \rangle,\\
 \Gamma^h_{ij,mn} &\equiv \Gamma^p_{in,jm},
\end{split}
 {\label{eq:bare}}
\end{equation}
and they are diagrammatically shown in Fig.~\ref{phi} (a) and (b).
The Hartree-Fock (HF) and the second order $\Phi$ diagrams can be
drawn easily by the bare vertices [Fig.~\ref{phi} (c) and (d)].
Connecting the bare PP or PH vertices in the
random-phase-approximation (RPA) series by the Green's function
$\hat G$, we get the main body of FLEX $\Phi$ diagrams, which is shown in
Fig.~\ref{phi} (e) and (f).

For the $m$-orbital system without  SU(2) symmetry, the
Green's function and the self-energy can be expressed as $2m\times
2m$ matrices, satisfying the Dyson equation,
\begin{equation}
\hat G(k,ik_n)= [ik_n \hat1- \hat{h}(\bm k) -\hat\Sigma(\bm k,ik_n)]^{-1},
{\label{eq:Dyson}}
\end{equation}
where $\hat{h}(\bm k)$ represents the free part Hamiltonian
including SOC [see Eq.~(\ref{eq:H0})]. The self-energy is obtained
by taking the derivative of $\Phi$ with respect to $\hat G$, i.e.,
plucking one line of Fig.~\ref{phi} (c) to (f),
\begin{equation}
\begin{split}
  \Sigma_{ij}(k) =\ & \frac{T}{N} \sum_{q;mn}\{\Gamma^h_{ij,mn}G_{mn}(k\!-\!k')\text{e}^{i(k_l\!-\!k'_l)0^+}\\
                 &+0.5[\hat\Gamma^h\hat\chi^{h0}(q)\hat\Gamma^h]_{im,jn}G_{mn}(k\!-\!q)\\
                 &-[\hat\Gamma^p(\hat\chi^{p}(q)-\hat\chi^{p0}(q))\hat\Gamma^p]_{in,jm}G_{mn}(q\!-\!k)\\
                 &+[\hat\Gamma^h(\hat\chi^{h}(q)-\hat\chi^{h0}(q))\hat\Gamma^h]_{im,jn}G_{mn}(k\!-\!q)\},
\end{split}
{\label{eq:self}}
\end{equation}
where the  abbreviation $k\equiv[\bm k, ik_l]$ ($q\equiv[\bm q,
iq_l]$) is used with the fermion (boson) Matsubara frequency
$k_l=(2l+1)\pi T$ ($q_l=2l\pi T$). The
first and second terms in Eq.~(\ref{eq:self}) represent the HF and
second order contributions,  and the third and forth terms
represent the RPA-bubbles of the PP and PH channels. The
susceptibilities for these two channels are given by the
$(2m)^{2}\times (2m)^{2}$ matrices,
\begin{equation}
\begin{split}
\hat\chi^{p}(q)&=\ \hat\chi^{p0}(q)[\hat 1 + \hat\Gamma^p\hat\chi^{p0}(q)]^{-1},\\
\hat\chi^{h}(q)&=\ \hat\chi^{h0}(q)[\hat 1 + \hat\Gamma^h\hat\chi^{h0}(q)]^{-1},
\end{split}
{\label{eq:chi}}
\end{equation}
in which the Linhard functions  are  defined respectively by,
\begin{equation}
\begin{split}
\chi^{p0}_{ij,mn}(q)&= \frac{T}{2N}\sum_{k}G_{im}(q\!-\!k)G_{jn}(k),\\
\chi^{h0}_{ij,mn}(q)&= -\frac{T}{N}\sum_{k}G_{im}(k\!+\!q)G_{nj}(k).
\end{split}
{\label{eq:chi0}}
\end{equation}
Equations (\ref{eq:Dyson}) to (\ref{eq:chi0}) form a closed set
and thus could be solved self-consistently.

The irreducible vertex of the Bethe-Salpeter (BS) equation for the
PH channel is the differentiation of self-energy (or the second
derivative of $\Phi$ ) with respect to $\hat G$.  Apparently there
are  two kinds of contributions: (a) the diagrams in which the
plucked two lines belong to the same bubble and only single
RPA-form fluctuation is included. (b) Aslamazov-Larkin (AL)
diagrams in which the plucked two lines belong to different
bubbles and more than one RPA-form fluctuations are taken into
account. Here we omit the AL diagrams for the following
reasons~\cite{flex-1,flex-3,Bickers_book}: First, the AL contributions
are demonstrated to be small (no more than $15\%$ for the
zero-momentum static susceptibility, for example). Second, the AL
contributions will aggravate the agreement between the FLEX
results and the benchmark QMC simulations. Third, the AL diagrams
are necessary  if we require keeping the conservation law
rigorously, which is believed to be essential in some transport
studies\cite{Kadanoff_book}. Since here we investigate the
equilibrium properties of electrons in the normal state, the AL diagrams can be safely removed.
Therefore we obtain the PH channel's irreducible vertex
\begin{equation}
\begin{split}
I^h_{ij,mn}(k,k';Q) =\ & \Gamma^h_{ij,mn} - [\hat\Gamma^h\hat\chi^h(k\!-\!k')\hat\Gamma^h]_{im,jn} \\
                    &+ [\hat\Gamma^p\hat\chi^p(k\!+\!k'\!+\!Q)\hat\Gamma^p]_{in,jm},
\end{split}
{\label{eq:Ih}}
\end{equation}
where  momenta $k$, $k'$ are for the fermions and  $Q$ is for the collective
bosonic  mode. The corresponding BS equation reads
\begin{equation}
\begin{split}
\sum_{k';\alpha\beta}I^h_{ij,\alpha\beta}(k,k';Q)G_{\alpha m}(k'\!+\!Q)G_{n\beta}(k')\psi^{h}_{mn}(k';Q)\\
=\lambda^h_Q\psi^{h}_{ij}(k;Q),
\end{split}
{\label{eq:BSh}}
\end{equation}
where $\lambda^h_Q$ and $\hat\psi^h(k;Q)$ represent the eigenvalue
and eigenfunction. Particularly, if  $\lambda^h_Q$ approaches to
$1$  at zero frequency, the system undergoes  a spontaneously
symmetry breaking at momentum $\bm Q$ in the PH channel.

The  irreducible vertex for PP channel is the second derivative of
$\Phi$  with respect to anomalous Green's functions $\hat F$ and
$\hat F^\dag$~\cite{Tewordt}, here $F_{ij}=\langle T_\tau c_i
c_j\rangle$ and $F_{ij}^\dag=\langle T_\tau
c^\dag_ic^\dag_j\rangle$. Although there is no anomalous Green's
function lines in the original $\Phi$ diagrams (see Fig.~\ref{phi}), we
can construct such diagrams by replacing two $\hat G$  with $\hat
F$ and $\hat F^\dag$  but keeping two arrows pointing in and other
two arrows pointing out at each dot. Here we omit the AL diagrams
again and obtain the PP channel's irreducible vertex
\begin{equation}
I^p_{ij,mn}(k,k';Q) = \frac{1}{2}\Gamma^p_{ij,mn} - [\hat\Gamma^h\hat\chi^h(k-k')\hat\Gamma^h]_{im,nj},
{\label{eq:Ip}}
\end{equation}
and the  corresponding BS  equation
\begin{equation}
\begin{split}
-\sum_{k';\alpha\beta}I^p_{ij,\alpha\beta}(k,k';Q)G_{\alpha m}(k'\!+\!Q)G_{\beta n}(-k')\psi^{p}_{mn}(k';Q)\\
=\lambda^p_Q\psi^p_{ij}(k;Q).
\end{split}
{\label{eq:BSp}}
\end{equation}
Unlike the PH channel, the largest eigenvalue $\lambda^p_Q$ always
associates with  $Q=[\bm 0,0]$, indicating the formation of Cooper
pairs with opposite momenta. Eqs.~(\ref{eq:Ip}) and ~(\ref{eq:BSp}) can
be used to investigate the most favorable superconducting pairing
gap which corresponds to $\hat\psi^p(k;Q)$ with the largest value
of $\lambda^p_Q$.

We have completed the introduction of the formal FLEX
formulations, in practical application some reasonable
approximations are widely used to simplify  the computations.
First, when the on-site interactions are all repulsive, the
contributions of the PP RPA-bubbles in Fig. \ref{phi} (e) are
relatively small and therefore could be safely left out
~\cite{flex-1,flex-3,Takimoto,Note_PP}. Second, it is more
convenient to choose other numerical criterions, instead of
Eq.~(\ref{eq:BSh}), to evaluate the PH channel's instability. For
example, one can choose the Stoner-like criterion
$\text{det}[\hat1+ \hat\Gamma^h\hat\chi^{h0}(\bm
Q,0)]<0.002$~\cite{Takimoto}, or if the biggest element of
$\chi^h_{ij}(\bm  Q,0)$ is 50 times larger than
$\chi^{h0}_{ij}(\bm Q,0)$~\cite{HuWang}, as we employ in this
paper.

To apply the above FLEX method to our study,
 the Hugenholtz vertices in Eq.~(\ref{eq:bare}) are
fabricated by the on-site interactions ($U,U,J,J^{\prime}$), and
the results presented in this paper are given by the parameters
($U,U,J,J')=(5,3.5,0.75,0.75$), no qualitative different results
are obtained when we change the values of $U$ and $J$. The
numerical calculations are performed on $64\times64$ $\bm k$
meshes with 1024 (for $T\!=\!0.04$), 2048 (for $T\!=\!0.02,
0.015$) and 4096 (for $T\!=\!0.01$) Matsubara frequencies.
Particularly, we utilize the technique developed by Deisz \emph{et
al.}~\cite{Deise} to efficiently include the contribution of high
Matsubara frequencies. The analytical continuation of Green's
functions to the real frequency is carried out by  Pad$\acute{\rm
e}$ approximation~\cite{pade}. The convergent solutions of the
FLEX equations are obtained  if the relative error of each matrix
element of $\hat \Sigma$ is smaller than $10^{-6}$.

\section{results and discussion}
\subsection{Collective excitations}

Let us first define corresponding susceptibilities for various collective excitations relevant to the following discussions.  The static transverse spin (TS) and
longitudinal spin (LS) susceptibilities are given by
\begin{equation}
\begin{split}
\chi^{\text{TS}}(\bm Q)=& \sum_{mn}\sum_{\alpha\beta\mu\nu}\sigma^x_{\alpha\beta}\sigma^x_{\mu\nu}
\chi^h_{m\beta \, m\alpha, n\mu  \, n\nu}(\bm Q,0), \\
\chi^{\text{LS}}(\bm Q)=& \sum_{mn}\sum_{\alpha\beta\mu\nu}\sigma^z_{\alpha\beta}\sigma^z_{\mu\nu}
\chi^h_{m\beta \, m\alpha, n\mu  \, n\nu}(\bm Q,0),
  \end{split}
  \label{spin_flu}
\end{equation}
where the spin and orbital degrees of freedom have been expressed by $(\alpha,\beta,\mu,\nu)$  and $(m,n)$, respectively. The charge fluctuation  is too small compared to other fluctuations and thus is not discussed here.
Because of the introduction of SOC, the contribution of orbital fluctuations is no longer neglectable, so we define the static transverse orbital (TO) and longitudinal orbital (LO) susceptibilities
\begin{equation}
\begin{split}
\chi^{\text{TO}}(\bm Q)=& \sum_{\alpha\beta}\sum_{ijmn}L^x_{ij}L^x_{mn}
\chi^h_{j\alpha \, i\alpha , m\beta  \, n\beta }(\bm Q,0), \\
\chi^{\text{LO}}(\bm Q)=& \sum_{\alpha\beta}\sum_{ijmn}L^z_{ij}L^z_{mn}
\chi^h_{j\alpha \, i\alpha , m\beta  \, n\beta }(\bm Q,0).
  \end{split}
  \label{orbital_flu}
\end{equation}

As discussed in the introduction, it has been suggested that the low-energy physics in Sr$_2$IrO$_4$ may be described as an effective one-band model with the isospin $J_{\text{eff}}=1/2$\cite{Kim1,science_kim}. To check its possible application here, we define the isospin susceptibility. The creation operators for the isospin-up and isospin-down states with momentum $\bm k$ are given by
$a^\dag_{\bm k,\Uparrow} = (d^\dag_{\bm k,1,\downarrow}+ i d^\dag_{\bm k,2,\downarrow}+ d^\dag_{\bm k,3,\uparrow} )/\sqrt{3}$ and
$a^\dag_{\bm k,\Downarrow} = (d^\dag_{\bm k,1,\uparrow}- i d^\dag_{\bm k,2,\uparrow} - d^\dag_{\bm k,3,\downarrow} )/\sqrt{3}$,
where indexes 1, 2, 3 denote $yz$, $zx$, and $xy$ orbitals.
Therefore, the isospin operator can be constructed:
$\bm{S}^\alpha_{\bm q}= \sum_{\bm k}(a^\dag_{\bm{k+q},\Uparrow},a^\dag_{\bm{k+q},\Downarrow})\bm{\sigma}^\alpha(a_{\bm{k},\Uparrow},a_{\bm{k},\Downarrow})^\text{T}$. Then, the static transverse isospin (TI) and
longitudinal isospin (LI) susceptibilities are given by,
\begin{equation}
\begin{split}
\chi^{\text{TI}}(\bm Q) = & \frac{T}{N}\int_{0}^{\frac{1}{T}}\langle T_\tau S^x(\bm q,\tau)S^x(-\bm q,0)\rangle d\tau,\\
\chi^{\text{LI}}(\bm Q) = & \frac{T}{N}\int_{0}^{\frac{1}{T}}\langle T_\tau S^z(\bm q,\tau)S^z(-\bm q,0)\rangle d\tau.
  \end{split}
  \label{iso_flu}
\end{equation}

\begin{figure}
  \centering
  \includegraphics[scale=1.5]{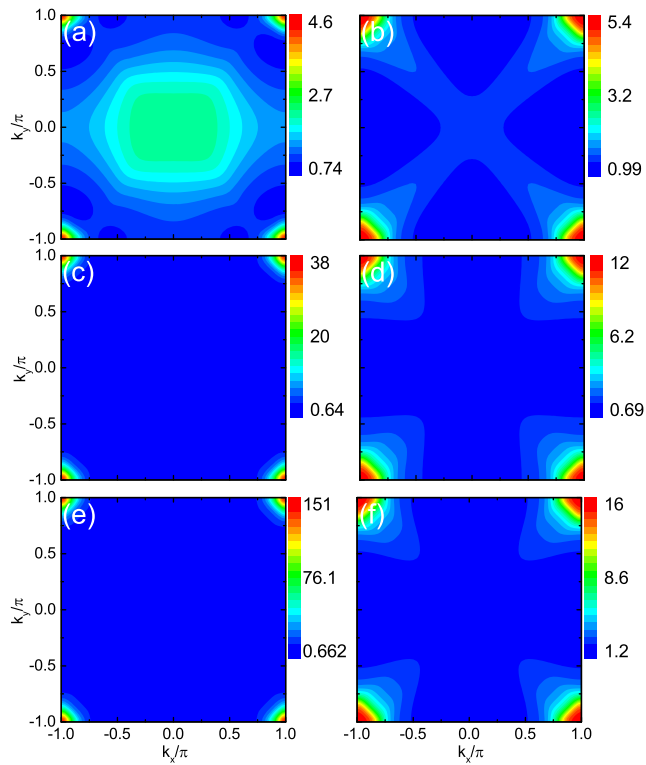}
  \caption{ (Color online) Susceptibility for various collective excitations at $3\%$ electron-doping and $T=0.01$.
   (a), (b), (c) and (d) show static TS,
  LS, TO  and LO  susceptibilities, respectively. (e) and (f) show static  TI and
  LI susceptibilities (see text). The TS fluctuation in (a) no longer exhibits the $C_4$
symmetry with the introduction of SOC.}
  {\label{sus1}}
\end{figure}

The static TS, LS, TO and LO susceptibilities for $3\%$
electron-doping at $T=0.01$ are shown in Fig.~\ref{sus1} (a)-(d).
As expected, the  strong peaks  exist around ($\pm\pi,\pm\pi$),
which are  reminiscences of the AFM order in the parent compound. In
particular, the intensities of orbital susceptibilities are all
stronger than those of spin susceptibilities, and the TO
fluctuation overwhelms the LO fluctuation. These features
are consistent with the experimental results in the single-crystal
neutron diffraction~\cite{finding2} and nonresonant magnetic X-ray
diffraction~\cite{L_larger_S}.

\begin{figure}
  \centering
  \includegraphics[scale=1.2]{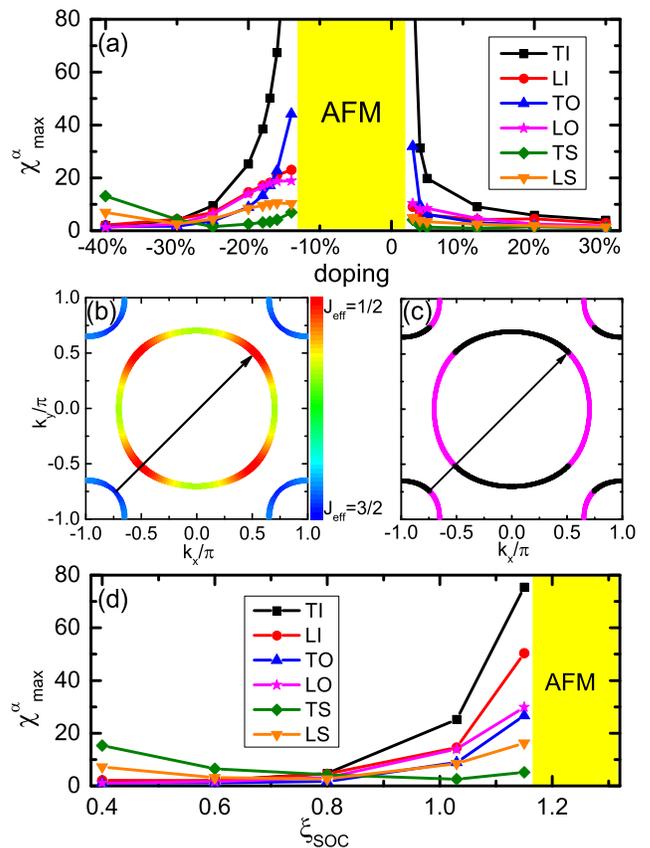}
  \caption{ (Color online)(a) The largest value of various susceptibilities $\chi^\alpha(\bm q)$ as a function of
  electron doping (positive) and hole doping (negative) at $T=0.02$, with $\alpha=$TI, LI, TO, LO, TS or LS.
  (b) Fermi surfaces for $40\%$ hole doping with the different colors representing the weights of the $J_{\text{eff}}=1/2$ doublet and $J_{\text{eff}}=3/2$ quartet. (c) The same as (b), but with the colors indicating the majority orbital character (magenta: $d_{xz}$, black: $d_{yz}$).
  (d) The largest value of $\chi^\alpha(\bm{q})$ as a function of the SOC magnitude $\xi_{SOC}$ at $T=0.02$ for $20\%$ hole doping. The lines with an arrow in (b) and (c) indicate the nesting vector.
  }
  \label{sus0}
\end{figure}

Fig.~\ref{sus1} (e) and (f) show the TI and LI susceptibilities
for the same $3\%$ electron doping at $T=0.01$. The peaks also
reside at the same momenta as in the spin and orbital
fluctuations. Quite strikingly, the intensity of the TI
susceptibility  is almost one order of magnitude larger than that
of all other fluctuations. To show in more detail the effects of
the isospin fluctuations, in Fig.~\ref{sus0} (a) we compare the
maximum magnitude of $\chi^\alpha(\bm q)$ at $T=0.02$, from 40\%
hole-doping to 30\% electron-doping, with $\alpha$ denoting TI,
LI, TO, LO, TS and LS channels. The AFM order is determined by the
numerical criterion in the PH channel, as introduced in Sec.II B.
As shown, the TI susceptibility tends to diverge as the system
approaches the AFM state, from both the electron-doped side and the
hole-doped side. These results suggest that the AFM order in the
undoped and slightly doped system is mainly caused by the TI
fluctuation. Consequently, it shows that the effective one-band
model with the isospin
$J_{\text{eff}}=1/2$~\cite{Kim1,science_kim} can describe
 the magnetic properties reasonably in the undoped and slightly
doped Sr$_2$IrO$_4$.
Another  feature drawn from Fig.~\ref{sus0} (a) is that the AFM is
more robust against the hole-doping in comparison to the electron-doping,
which may be useful for the related experimental investigations.

When the system is doped away from the AFM region, Fig.~\ref{sus0}
(a) shows that the TI fluctuation decreases rapidly with doping.
However, it still has larger magnitude than all others in the
region of 30\% hole-doping to 50\% electron-doping (the results
for $>30\%$ electron-doping are not shown here), though their
differences are decreased with dopings. In this  region, the
momentum $\bf q$ is at or  near the $(\pi,\pi)$ point. When doping
the system further with holes ($>30\%$), the spin fluctuation
becomes dominant so that the effective one-band picture is broken
down. These results can be understood from the weights of the
$J_{\text{eff}}=1/2$ doublet and $J_{\text{eff}}=3/2$ quartet
along the Fermi surface, as shown in Fig.~\ref{sus0} (b).
One can see that the closed electron pocket centered around the
$\Gamma=(0,0)$ point is  composed mainly of the
$J_{\text{eff}}=1/2$ doublet, while the hole pocket around
$(\pi,\pi)$ mainly of the $J_{\text{eff}}=3/2$ quartet. For $40\%$
hole-doping, the hole Fermi pocket becomes large and introduces a Fermi surface nesting between the hole and electron pockets, as
shown in Fig.~\ref{sus0} (b). Thus, the inter-pocket particle-hole
scattering between the $J_{\text{eff}}=1/2$ dominant band and
$J_{\text{eff}}=3/2$ dominant band takes action and overwhelms
the intra-pocket scattering in the $J_{\text{eff}}=1/2$ dominant band,
which makes the isospin fluctuation no longer the leading collective excitation.
On the other hand, from the distribution of orbital characters shown in Fig.~\ref{sus0} (c),
we can find that the main inter-pocket scattering occurs in the same orbital, which makes
the orbital fluctuations also suppressed. Therefore, the spin fluctuation is the dominant collective excitation
for the heavily hole-doped system.
However, the Fermi level moves away from
the $J_{\text{eff}}=3/2$ dominant band with the electron doping, consequently it is only the
$J_{\text{eff}}=1/2$ dominant band that crosses the Fermi level.
Therefore, the one-band picture is
always robust against the electron doping. Furthermore, we have
also examined  the range of validity of the effective one-band
picture with respect to the strength of SOC. As shown in
Fig.~\ref{sus0} (d), the TI susceptibility prevails all others for
20\% hole-doping if  $\xi_{SOC}>0.8$, indicating that this picture
survives in an extended range.

\subsection{Weak pseudogap behavior and Fermi arcs}
\begin{figure}
  \centering
  \includegraphics[scale=0.7]{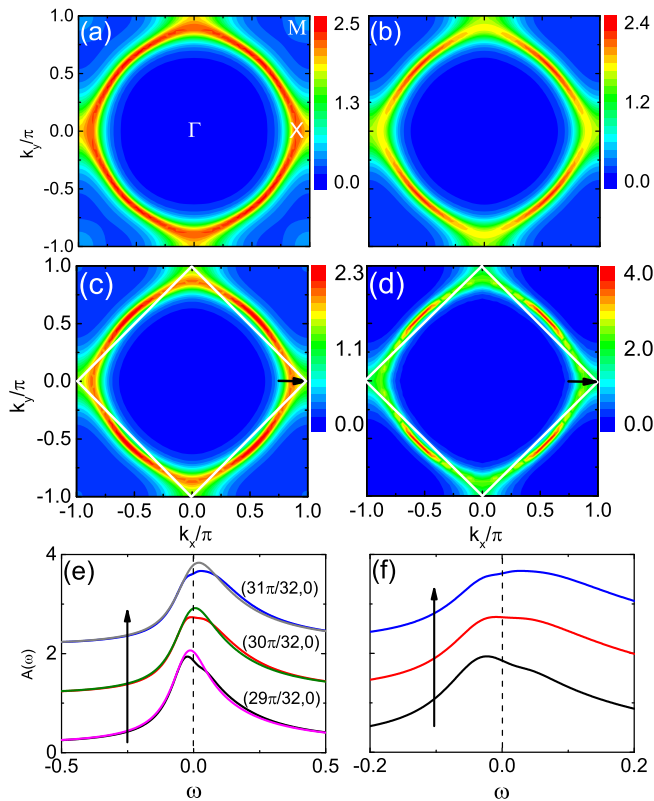}
  \caption{(Color online) Spectral functions $A(\bm k,0)$ for 10\% electron-doping (a) and 5\% electron-doping (b) at  $T=0.015$. Spectral functions $A(\bm k,0)$ at $T=0.02$ (c) and  $T=0.01$ (d) for 3\% electron-doping.
  The pink (black), green (red), gray (blue) lines in (e) are $\omega$ dependence of the spectral functions
  for $T=0.02$ ($T=0.01$), at three momenta  indicated by the arrow in (c) and (d).
  An enlarged version for $T=0.01$ is shown in (f). The white lines in (c) and (d) denote the boundary of the magnetic Brillouin
zone.}
  {\label{Gk_nT}}
\end{figure}
The pseudogap behavior can be detected from the
single particle spectral function which is defined as $A(\bm
k, \omega)= -\sum_{n}\text{Im}G_{nn}(\bm k,\omega)/\pi$, where $n$
denotes both spin and orbital indices. In Fig.~\ref{Gk_nT} (a) and
(b), we show the contour plot of the zero-energy spectrum at
$T=0.015$ for $10\%$ and $5\%$ electron-doping, respectively.
Since the system has been  away from the AFM  order at these dopings [see Fig.~\ref{sus0} (a)],
an intact Fermi
surface is expected in the conventional normal state. Indeed, for
$10\%$ electron-doping, a closed diamond-shaped Fermi surface is
observed, as shown in Fig.~\ref{Gk_nT} (a). However, an obvious
reduction in the spectral intensity occurs around $(0,\pi)$ and its symmetric points for
$5\%$ electron-doping. This reduction is strongly
temperature-dependent, because it appears only below a certain
temperature, as shown in Fig.~\ref{Gk_nT} (c) and (d) at $3\%$
electron-doping for $T=0.02$ and $T=0.01$, respectively.
Owing to the suppression of the spectra near $(0,\pm \pi)$ and
$(\pm \pi,0)$, the Fermi surface around these momenta is
destructed and the four residual separated segments
form the so-called Fermi arcs [see Fig.~\ref{Gk_nT} (b) and (d)].
These results are consistent with the recent ARPES
experiment~\cite{arc1}. In order to look in more detail the
suppression of the spectra, we show the energy distribution curves
(EDC) of the spectral functions for $3\%$ electron-doping at three
momentum points near $(\pi,0)$ in Fig.~\ref{Gk_nT} (e). One can
see that the suppression occurs basically around the Fermi energy
when the temperature is decreased from $T=0.02$ to $T=0.01$.
Moreover, different from a well defined quasiparticle peak at
$T=0.02$, the spectral functions at $T=0.01$ for ${\bf
q}=(30\pi/32,0)$ and $(31\pi/32,0)$ show a weak dip around the
Fermi energy, as shown in Fig.~\ref{Gk_nT} (f). It suggests that a weak pseudogap
does exist around $(\pi,0)$  and its symmetric momenta for
slightly electron-doping.

Since thers is no long-range order in this doping range,
the pseudogap is most likely to  result from
the scattering of quasiparticles by the collective excitations. In
this framework, the quasiparticles around $(\pi,0)$ and $(0,\pi)$
are strongly coupled by the collective excitations with the transferred momentum
$(\pi,\pi)$  in the
scattering process. In accord with this analysis, the results
presented in Fig.~\ref{sus1} demonstrate that all collective
excitations, including the spin, orbital and isospin fluctuations,
exhibit the characteristic momentum $(\pi,\pi)$. In particular,
the transverse isospin (TI) fluctuation overwhelms the spin and
orbital fluctuations. We have also carried out a careful
examination, and find that all the important elements of
$\hat\chi^h(\bm q,0)$ are included in the TI fluctuation.
Especially, for ${\bm q}=(\pi,\pi)$ the elements included in the
TI susceptibility are calculated to amount to 91.3\% weight of
the total PH susceptibilities for 3\% electron-doping at $T=0.01$.
Therefore, it suggests that the scatterings of quasiprticles by
the ($\pi,\pi$) TI fluctuation lead to a weak pesudogap partially
opened around $(0,\pm\pi)$ and $(\pm\pi,0)$.

\begin{figure}
  \centering
  \includegraphics[scale=0.82]{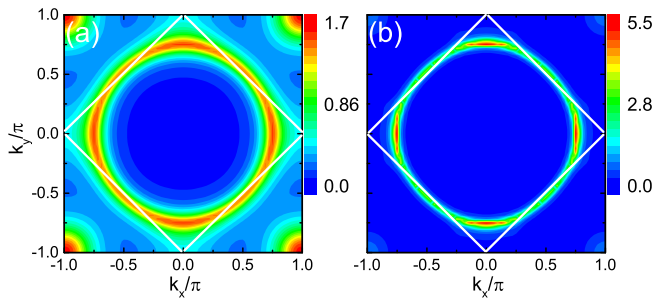}
  \caption{ (Color online)
  (a) and (b) are spectral functions $A(\bm k,0)$ at  $T=0.04$  and $T=0.01$  for 17\% hole-doping.
 The white lines denote the boundary of the magnetic Brillouin zone.}
  {\label{FA_hole}}
\end{figure}

We have also investigated the possible pseudogap behavior and
Fermi arcs in the lightly hole-doped region. The typical results
of the spectral function $A(\bm k,0)$ for 17\% hole-doping are
shown in Fig.~\ref{FA_hole}. At $T=0.04$, there is a nearly
circular Fermi surface around the $\Gamma$ point. By decreasing
the temperature to $T=0.01$, the spectral weights are suppressed
on some Fermi momenta  and consequently it leads to
the formation of the Fermi arcs. However, the suppressions now
appear around $(\pi/2,\pi/2)$, which is contrary to the case of
the electron-doping shown in Fig.~\ref{Gk_nT} (b) and (d).
This suppression  also results from the ($\pi,\pi$) TI
fluctuation. In this case, the Fermi surface shrinks in comparison
with that of electron-doping, so that the ``hot spots" (the
crossing points of the Fermi surface with the boundary of the AFM Brillouin zone)
at which the quasiprticles are strongly
scattered move from the $(\pm\pi,0)$ and $(0,\pm\pi)$ points to the $(\pm\pi/2,\pm\pi/2)$ points
(see Fig.~\ref{Gk_nT} and Fig.~\ref{FA_hole}).

\section{Summary}
In summary, we have extended the FLEX  approach by Hugenholtz
diagrams to include the case where  SU(2) symmetry is broken.
Using this approach, we investigate various collective
fluctuations and the spectral function of quasiparticles. It is
found that the isospin fluctuations derived from the
$J_{\text{eff}}=1/2$ spin-orbit Mott insulator dominate over the
spin, orbital and charge fluctuations in the extended doping
region, suggesting the validity of the isospin
$J_{\text{eff}}=1/2$ picture in an extended doping regime. Also
the isospin fluctuation leads to the emergence of the pseudogap
and Fermi arcs in the slightly doped system, which is consistent
with the ARPES experiments for slightly doped
Sr$_2$IrO$_4$~\cite{arc1,YCao}.

\begin{acknowledgments}
We would like to acknowledge Q. H. Wang  and   H. Y. Zhang for
discussions on the generalized FLEX method,  H. Yao and J. Kang
for their useful discussions. This work was supported by the
National Natural Science Foundation of China (11190023, 11204125
and 11404163), the Ministry of Science and Technology of China
(973 Project Grants No.2011CB922101 and No. 2011CB605902).
\end{acknowledgments}


\begin{references}
\bibitem{Note_PP} Indeed, we have checked that our calculations show no qualitative differences with or without PP RPA-bubbles.
\end{references}
\end{document}